\documentclass[conference]{IEEEtran}
\IEEEoverridecommandlockouts
\usepackage{cite}
\usepackage{amsmath,amssymb,amsfonts}
\usepackage{algorithmic}
\usepackage{graphicx}
\usepackage{textcomp}
\usepackage{xcolor}
\def\BibTeX{{\rm B\kern-.05em{\sc i\kern-.025em b}\kern-.08em
    T\kern-.1667em\lower.7ex\hbox{E}\kern-.125emX}}

\usepackage{amsmath}
\usepackage{amsmath, bm}
\usepackage{amsfonts}
\usepackage{amsthm}
\usepackage{url}
\usepackage{bm}
\usepackage{mathrsfs}

\theoremstyle{definition}

\theoremstyle{remark}

\usepackage{algorithm}
\usepackage{algorithmic} 

\usepackage{enumerate}

\usepackage{multirow}
\usepackage{multicol}
\usepackage{booktabs}
\usepackage{makecell}
\usepackage{threeparttable}
\usepackage{array}
\usepackage{footnote}

\usepackage{xcolor}

\begin{document}
\title{Transmission Congestion Management with Generalized Generation Shift Distribution Factors
\thanks{The work was supported by the Science and Technology Project of State Grid Corporation of China 5108-202155045A-0-0-00.}}

\author{
    \IEEEauthorblockN{
        Shutong Pu,
        Guangchun Ruan,
        Xinfei Yan,
        and Haiwang Zhong}
    \IEEEauthorblockA{State Key Lab of Power Systems, Department of Electrical Engineering
    \\Tsinghua University, Beijing 100084, China\\Email: zhonghw@mail.tsinghua.edu.cn}
}
\maketitle


\begin{abstract}

A major concern in modern power systems is that the popularity and fluctuating characteristics of renewable energy may cause more and more transmission congestion events.
Traditional congestion management modeling involves AC or DC power flow equations, while the former equation always accompanies great amount of computation, and the latter cannot consider voltage amplitude and reactive power.
Therefore, this paper proposes a congestion management approach incorporating a specially-designed  generalized generator shift distribution factor (GSDF) to derive a computationally-efficient and accurate management strategies. 
This congestion management strategy involves multiple balancing generators for generation shift operation. The proposed model is superior in a low computational complexity (linear equation) and versatile modeling representation with full consideration of voltage amplitude and reactive power.
During the congestion management process, the generalized GSDF instructs the selection of target generators while linearized AC equation reduces the modeling complexity.   
We use the IEEE 118-bus system to validate the proposed model and approaches, whose outcomes are found to be more accurate than the baseline sensitivities calculated by DC power flow.

 \end{abstract}


%
\IEEEpeerreviewmaketitle

\section{Introduction}

Towards the future renewable-dominated power systems, the transmission congestion management~\cite{4384982} will be of great importance and attract a public concern of operational security~\cite{2018ch}. Despite the regulatory role in marketing perspective, these management schemes avoid load shedding and stabilize transmission power flow.

Existing strategies in this field often consumed considerable amount of computational resources to make an optimal generation reschedule based on AC power flow~\cite{7407415}. For cases using mixed integer nonlinear programming approach, deriving the optimal strategies is a challenge because of the exponentially growing computing complexity~\cite{1709109, 9594777}. Therefore, existing works such as~\cite{gaonkar_nanannavar_manjunatha_2017} and \cite{4762626} tried using sensitivity analysis (based on DC power flow) and introducing indicators for transmission congestion management. Optimal power flow model was fundamental for congestion management as denoted in \cite{7079504}, and some tailored control models were developed to incentive diversified resources to secure the whole system~\cite{2020con}. However, these works are mostly based on DC power flow which ignores reactive power flow and assumes constant voltage amplitude.

In this paper, we apply an advanced linear approximation model for AC power flow~\cite{8468081}, which is able to simplify the congestion management modeling and consider the factors of voltage amplitude and reactive power with good accuracy. Based on this power flow model, we first generalize the generation shift distribution factor to consider the full influential factors including reactive power and voltage amplitude. Meanwhile, congestion release operation should also prevent re-congestion in neighbouring transmission lines. For efficiency and feasibility purposes, a novel congestion management scheme using changeable balancing generators is then developed as well.

In brief, this paper makes two contributions:
\begin{itemize}
    \item It applies a linearized AC power flow equation in congestion management with discernible benefits of accuracy and computing efficiency.
    \item It proposes generalized generator shift distribution factor for sensitivity analysis, and applies changeable balancing generators to improve the management efficiency.
\end{itemize}




\section{Generalized Generation Shift Distribution Factor (GSDF)}

This section will include the methological details of linearized (optimal) power flow, sensitivity analysis, and the generalized GSDF. A precision test is given at last.


\subsection{Linear Approximated Power Flow Equation}
Unlike AC power flow, the linearized power flow equation~\cite{7954975} contains linear functions with respect to voltage phase angle and square of voltage amplitude, formulated as:
\begin{align}
  &{P_{i,j}}={g_{ij}}\frac{v_i^2-v_j^2}{2}-{b_{ij}}{\theta_{ij}}+{{P_{i,j}}^{Loss}} \label{NCED:nonl} \\
  &{{P_{i,j}}^{Loss}}=g_{ij}[\frac{{{\theta_ij}^2}}{2}+\frac{1}{8}({v_i^2-v_j^2})^2] \label{NCED:nonloss}
\end{align}

\subsection{Optimal Power Flow and Sensitivity Analysis}
With the linearized power flow equation, we hereby establish an optimal power flow (OPF) model with the following objective function that minimizes the total cost of generators:
\begin{align}
  \min_{p_{i,t}} \sum_{i=1}^N \left( a_ip_{i,t}^2+b_ip_{i,t}+c_i \right) \label{NCED:obj}
\end{align}
where $N$ denotes the number of generators, and we consider a quadratic expression of the output $p_{i,t}$ to account the cost.

Power balance constraint ensures that the generation output meets demand, formulated as follows:
\begin{align}
  &\sum_{i=1}^N p_i = \sum_{i=1}^N L_i \quad i\in N\label{NCED:cons_demand}
\end{align}

For security reason, all the branch power flows should be enforced within a set of limits. These requirements are expressed as follows:
\begin{align}
p_i^{min} \leq p_i &\leq p_i^{max} \quad i=1,2,\dots,N_{gen} \label{NCED:cons_plimits}\\
q_i^{min} \leq q_i &\leq q_i^{max} \quad i=1,2,\dots,N_{gen} \label{NCED:cons_qlimits}\\
-T_k^{max} \leq T_k &\leq T_k^{max} \quad k=1,2, \dots , N_{br} \label{NCED:cons_Tlimits}
\end{align}

\subsection{Generalized GSDF}
Given the linearized power flow and sensitivity analysis results, we define the generalized GSDF as follows:
\begin{align}
  GSDF_{ij,k}=\frac{\partial{P_{ij}^{pf}}}{\partial{P_{G_k}^{trade}}}\label{NCED:GSDF1}
\end{align}
where $P_{ij}^{pf}$ is the power flow on line $ij$; And $\partial{P_{G_k}^{trade}}$ is the power generation transaction volume of the generator at node $k$ (that is, the change amount of active power injection at node $k$). This formula represents that if the output of the generator at node $k$ is fine-tuned, the line power flow must be changed accordingly, and the power flow of different lines is affected differently by the output of the generator at node $k$. GSDF represents the magnitude of such influence. Therefore, from the definition, GSDF can be calculated by changing the active output of the generator.

In particular, for AC power flow, generation shift of active ouput may alter both active power and reactive power on transmission lines. Here, it represents the shift distribution factor that only regulates the active power of  generation and  transmission.

Reactance matrix can directly calculate the generator shift distribution factor under DC power flow, as shown in the following function:
\begin{align}
    &GSDF_{ij,k}=\frac{\frac{\partial\theta_{ik}}{\partial{P_k}}-\frac{\partial{\theta_{jk}}}{\partial P_k}}{x_{ij}}=\frac{X_{ik}-X_{jk}}{x_{ij}}\label{NCED:GSDF2}
\end{align}

As for AC power flow, we apply the linear approximation mentioned above, and based on the chain rule of derivation we hereby propose the generalized GSDF:
\begin{align}
    &GSDF_{ij,k}=\frac{\partial{P_{ij}}}{\partial{P_{k}}}=\frac{\partial{P_{ij}}}{\partial{U_{ij}}}\cdot\frac{\partial{U_{ij}}}{\partial{P_{k}}}+\frac{\partial{P_{ij}}}{\partial{\theta_{ij}}}\cdot\frac{\partial{\theta_{ij}}}{\partial{P_{k}}}\label{NCED:GSDF3}
\end{align}
where $U_{ij}$ denotes the voltage square difference between node $i$ and node $j$:
\begin{align}
  &U_{ij}={v_i}^2-{v_j}^2\label{NCED:GSDF4}
\end{align}

To further calculate the generalized GSDF, we consider formula \eqref{NCED:nonl} without network loss terms:
\begin{align}
  &{P_{ij}}=\frac{1}{2}{g_{ij}}{U_{ij}}-{b_{ij}}{\theta_{ij}}\label{NCED:GSDF5}
\end{align}
where:
\begin{align}
  &\frac{\partial{P_{ij}}}{\partial{U_{ij}}}=\frac{1}{2}{g_{ij}},\frac{\partial{P_{ij}}}{\partial{\theta_{ij}}}=-{b_{ij}}\label{NCED:GSDF6}
\end{align}

Also we have the following equation according to formula \eqref{NCED:GSDF2}:
\begin{align}
    \frac{\partial{\theta_{ij}}}{\partial{P_{k}}}=\frac{\frac{\partial\theta_{ik}}{\partial{P_k}}-\frac{\partial{\theta_{jk}}}{\partial P_k}}{x_{ij}}=\frac{X_{ik}-X_{jk}}{x_{ij}}\label{NCED:GSDF7}
\end{align}
Therefore, the key of sensitivity deduction is to find derivative of $U_{ij}$. It is difficult to deduce the derivative of $U_{ij}$ by theory, thus we introduce the optimal power flow model to simulate the derivative. 

In order to calculate the power transmission shift factor through simulation, the balancing generator needs to be chosen to meet the power generation and consumption balance equation\eqref{NCED:cons_demand}.

However, the balancing generator should be picked with caution. The generalized GSDF will change with the balancing generator, because every time the target generator increases one unit output, the balancing generator decreases one unit output. Therefore, the generalized GSDF actually represents the sensitivity of line power flow when the output of the balancing generator is transferred by one unit to the target generator (or the other way around). 

We will further discuss the selection of balancing generator in the next section.

In particular, since the objective function in the optimal power flow model is the lowest cost of the generator's active output, as shown in formula \eqref{NCED:obj}, the power flow change caused by the slight increase of the node's active power injection cannot be directly used for the calculation of GSDF.

Therefore, when increasing the active power injection of a generator node to calculate the generation shift distribution factor between target generator and the balancing generator, it is necessary to anchor the power injection of other nodes: including both active power injection and reactive power injection.
\begin{align}
    &P_i^{inj,state1}\leq P_i^{inj,state2}
    \leq P_i^{inj,state1}+\epsilon \quad i\neq k
    \label{NCED:GSDF_con1}\\
    &P_i^{inj,state2} = P_i^{inj,state1}+0.1MW \quad    i=k
    \label{NCED:GSDF_con2}\\
    &Q_i^{inj,state1}\leq Q_i^{inj,state2}
    \leq Q_i^{inj,state1}+\epsilon \quad    i\neq k
    \label{NCED:GSDF_con3}
\end{align}

Formula \eqref{NCED:GSDF_con1} indicates that the injected active power of node $i$ before and after the generator output disturbance added to node $k$ is almost unchanged ($P_i^{inj,state1}$). $\epsilon$ denotes the tolerance range which  is far less than $0.1 p.u.$
\subsection{Precision Test}

This test is focused on the precision of the proposed generalized GSDF. Ideally, the result shall be closer to AC modeling sensitivity than DC modeling sensitivity. We validate this statement in IEEE-9 bus system.


Parameter of generators are shown in Table~\ref{table:case_9gen}.

\newcommand{\tabincell}[2]{\begin{tabular}{@{}#1@{}}#2\end{tabular}}
\begin{table}
  \centering
  \caption{Parameter of Generators in the IEEE 9-bus System}
  \setlength\abovecaptionskip{-0.5cm}
  \setlength\belowcaptionskip{-0.5cm}
  \label{table:case_9gen}
  \renewcommand{\arraystretch}{1.3}
  \begin{threeparttable}
  \setlength{\tabcolsep}{2mm}{
    \begin{tabular}{ccccc}
      \toprule
      {Gen.}& {Capacity(MW)}& {Bus} &\tabincell{c}{Quadratic Cost \\Term($\$/{\text{MW}}^2$)} & \tabincell{c}{Primary Cost\\Term($\$/\text{MW}$)}\\ 
      \midrule
       G1 & 500 & 1 & 0.1100 & 5.0\\
       G2 & 590 & 2  & 0.0850 & 1.2\\
       G3 & 400 & 3  & 0.1225 & 1.0\\
        \bottomrule
    \end{tabular}
  }
  \end{threeparttable}
\end{table}

Parameter of transmission lines are shown in Table~\ref{table:case_9branch}.
\begin{table}[]
  \centering
  \caption{Parameter of Transmission Lines in the IEEE 9-bus System}
  \setlength\abovecaptionskip{-0.5cm}
  \setlength\belowcaptionskip{-0.5cm}
  \label{table:case_9branch}
  \renewcommand{\arraystretch}{1.3}
  \begin{threeparttable}
  \setlength{\tabcolsep}{3mm}{
    \begin{tabular}{cccc}
      \toprule
    \tabincell{c} {Transmission\\line}& \tabincell{c}{Start bus \& \\ end bus}  &{Capacity(MW)}&{Reactance($p.u.$)} \\ 
\midrule
       1 & 1-4 & 250 & 0.0576\\
       2 & 4-5 & 250  & 0.0920\\
       3 & 5-6 & 150  & 0.1700\\
       4 & 3-6 & 300 & 0.0586\\
       5 & 6-7 & 150  & 0.1008\\
       6 & 7-8 & 250  & 0.0720\\
       7 & 8-2 & 250 & 0.0625\\
       8 & 8-9 & 250  & 0.1610\\
       9 & 9-4 & 250  & 0.0850\\
        \bottomrule
    \end{tabular}
  }
  \end{threeparttable}
\end{table}

We select $Gen.2$ as target generator and $Gen.1$ as balancing generator. Then generalized GSDF denotes  the sensitivity of transmission line power flow when the output of  $Gen.1$ is transferred by 0.1~MW to $Gen.2$. Respectively apply DC power flow equation, AC power flow equation and the linear approximation of AC power flow equation into OPF model, and calculate $GSDF_{G_2}$ under constraints \eqref{NCED:GSDF_con1}, \eqref{NCED:GSDF_con2} and  \eqref{NCED:GSDF_con3}. The results are as follows:

\begin{table}[]
  \centering
  \caption{Comparison of Precision Result in the IEEE 9-bus System}
  \setlength\abovecaptionskip{-0.5cm}
  \setlength\belowcaptionskip{-0.5cm}
  \label{table:case_9result}
  \renewcommand{\arraystretch}{1.3}
  \begin{threeparttable}
  \setlength{\tabcolsep}{3mm}{
    \begin{tabular}{cccc}
      \toprule
     \tabincell{c} {Transmission\\line} &\tabincell{c} {DC modeling\\ GSDF} &\tabincell{c}{Approximated AC \\modeling GSDF}&\tabincell{c}{AC modeling \\GSDF} \\ 
\midrule
       1 & 1.0000 & 1.0000 & 0.9679\\
       2 & 0.3613 & 0.3524  & 0.3486\\
       3 & 0.3613 & 0.3624  & 0.3621\\
       4 & 0.0000 & 0.0000 & 0.0000\\
       5 & 0.3613 & 0.3580  & 0.3596\\
       6 & 0.3613 & 0.3612 & 0.3628\\
       7 & 1.0000 & 1.0000 & 0.9994\\
       8 & -0.6387 & -0.6135  & -0.6112\\
       9 & -0.6387 & -0.6179 & -0.6161\\
        \bottomrule
    \end{tabular}
  }
  \end{threeparttable}
\end{table}

\begin{figure}[]
  \centering
  \includegraphics[width=0.48\textwidth]{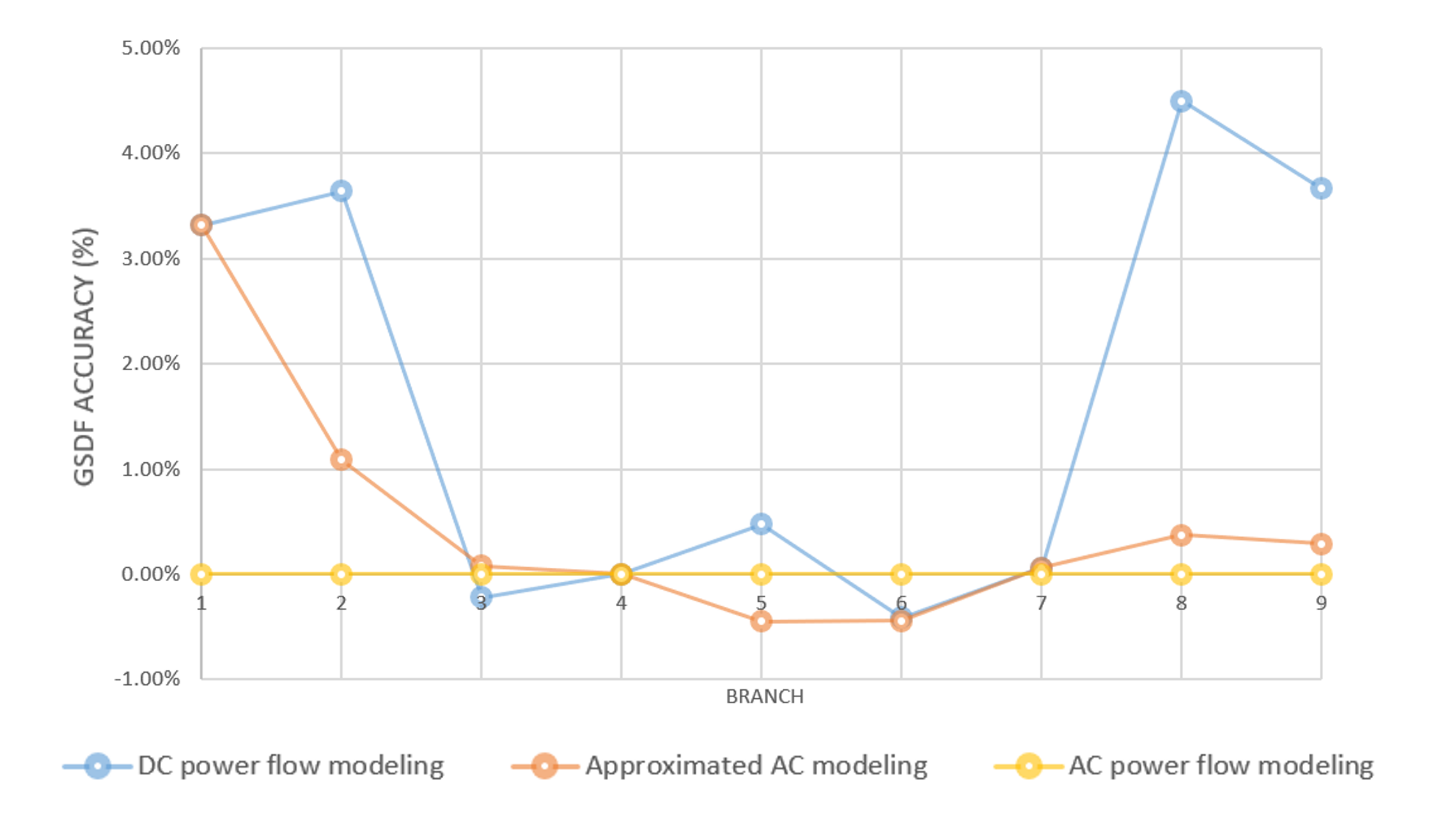} 
  \caption{Comparison of Precision Result in the IEEE 9-bus System} 
  \label{fig:1} 
\end{figure}

The accuracy of generalized GSDF  by approximated AC modeling is shown in Fig.~\ref{fig:1}. Taking AC power flow as the benchmark, the proposed generalized GSDF by linear approximation turns out to be more accurate than DC modeling result. However, the approximated AC method does not include extreme non-linear terms such as sine and cosine terms as in AC power flow equations.

Fig.~\ref{fig:1} and Table~\ref{table:case_9result} denotes that the proposed calculation of GSDF comes closer to AC result than to DC. 

This result shows that generalized GSDF is computational simpler than AC modeling, and meanwhile more accurate than traditional GSDF in DC modeling. This may increase the accuracy and efficiency of congestion management. We will further quantify the effect of this proposed GSDF in case study.

\section{Congestion Management Strategy}

Under the definition of generalized GSDF, this paper applies a congestion management strategy that involves multi balancing generators. The advantage of applying multi balancing generators compared to one-for-one match of targeting and balancing generator are listed as follows:

\begin{itemize}
\item changeable balancing generators lead to multiple generation shift, therefore multiple transmission shift. However single transmission shift with one-for-one match of targeting and balancing generators takes more risks of re-congestion.
\item changeable balancing generators result in less volatility of power flow in the system perspective, and balancing generators share the generation shift within feasible amount.
\end{itemize}

When a transmission line requires congestion management, in order to maintain constraints within limits and system operating normally, we follow the following principles during congestion management:

\begin{itemize}
\item To ensure efficiency, select the target generator which has the most significant value of generalized GSDF if the transmission line violates the power flow constraint;

\item Select one generator as balancing generator during each congestion management operation, and this generator should increase the exact output that the target generator decreases;

\item The balancing generator should be electrically distant from the target generator, so that the increase output of the balancing generator will not cause extra line congestion.

\item If ideal, the power flow change of the congested transmission line before and after the congestion management should be controllable.
\end{itemize}
Those principles assures the precision and efficiency of the congestion management strategy. According to those principles, we first calculate the generalized GSDF of the congested transmission lines. As mentioned in principle \textit{\textbf{c.}} ,  every time the target generator increases one unit output, the balancing generator decreases one unit output, thus GSDF changes with the selection of target generator and the balancing generator.

\begin{figure}[]
  \centering
  \includegraphics[width=0.32\textwidth]{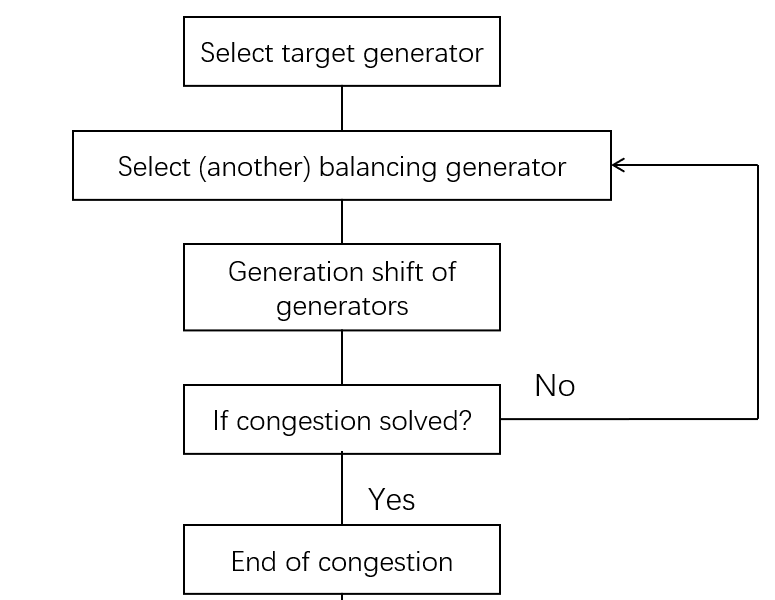} 
  \caption{Flowchart of the proposed congestion management strategy} 
  \label{fig:GSDF_trans} 
\end{figure}

Fig.\ref{fig:GSDF_trans} shows the relationship of generalized GSDF with different balancing generators. We may present those two GSDFs as $GSDF_{A}$ and $GSDF_{B}$, where A and B denotes different balancing generators. 

The generalized GSDFs of changeable balancing generators can be derived using the following formulas:
\begin{align}
    GSDF_{A}=GSDF_{B}+GSDF_{A,B}\label{NCED:GSDF_multi}
\end{align}where $GSDF_{A,B}$ denotes the generalized shift distribution factor with generator A as target generator and generator B as balancing generator.

At last, the flowchart of the congestion management strategy is given in Fig.~\ref{fig:GSDF_trans}.

\section{Case Study}

This section will validate the proposed congestion management strategy in an IEEE 118-bus system.
 
 
Part of parameters of transmission lines are shown in Table~\ref{table:1}, and part of parameter of generators are shown in Table~\ref{table:case_118gen}. Data in the tables is relevant to the model for congestion management.

Transmission line $8$ is selected as the target congestion manage transmission line in the system, and according to the congestion management strategy displayed in Fig.~\ref{fig:GSDF_trans}, balancing generator should be selected to achieve generation shift and therefore congestion relieved.

\begin{table}[]
  \centering
  \caption{Parameter of Transmission lines in the 118-bus System}
  \setlength\abovecaptionskip{-0.5cm}
  \setlength\belowcaptionskip{-0.5cm}
  \label{table:1}
  \renewcommand{\arraystretch}{1.3}
  \begin{threeparttable}
  \setlength{\tabcolsep}{3mm}{
    \begin{tabular}{cccc}
      \toprule
      \tabincell{c} {Transmission\\line} & \tabincell{c}{Start bus \& \\ end bus}  &{Capacity(MW)}&{Reactance($p.u.$)} \\ 
\midrule
       3 & 4-5 & 800  & 0.1080\\
       4 & 3-5 & 700 & 0.0540\\
       5 & 5-6 & 1000  & 0.0208\\
       6 & 6-7 & 800 & 0.0305\\
       7 & 8-9 & 580 & 0.0267\\
       8 & 8-5 & 770  & 0.0322\\
       9 & 9-10 & 700  & 0.0688\\
        \bottomrule
    \end{tabular}
  }
  \end{threeparttable}
\end{table}

Transmission line $7$ has the most strict capacity boundary among neighbouring transmission lines. Thus, it presents a congesting trend in the 24-hour simulation. In the following study we shall focus on this exact transmission line.

\begin{table}[]
  \centering
  \caption{Parameter of Generators in the IEEE 9-bus System}
  \setlength\abovecaptionskip{-0.5cm}
  \setlength\belowcaptionskip{-0.5cm}
  \label{table:case_118gen}
  \renewcommand{\arraystretch}{1.3}
  \begin{threeparttable}
  \setlength{\tabcolsep}{2mm}{
    \begin{tabular}{ccccc}
      \toprule
      {Gen.}& {Capacity(MW)}& {Bus} &\tabincell{c}{Quadratic Cost \\Term($\$/{\text{MW}}^2$)} & \tabincell{c}{Primary Cost\\Term($\$/\text{MW}$)}\\ 
       \midrule
       G2 & 300 & 4 & 0.2400 & 3.0\\
       G3 & 100 & 6  & 0.2250 & 1.2\\
       G4 & 300 & 8  & 0.1250 & 1.0\\
       G5 & 550 & 10  & 0.1850 & 1.2\\
       G6 & 185 & 12  & 0.900 & 1.0\\
        \bottomrule
    \end{tabular}
  }
  \end{threeparttable}
\end{table}

\begin{figure}
  \centering
  \includegraphics[width=0.50\textwidth]{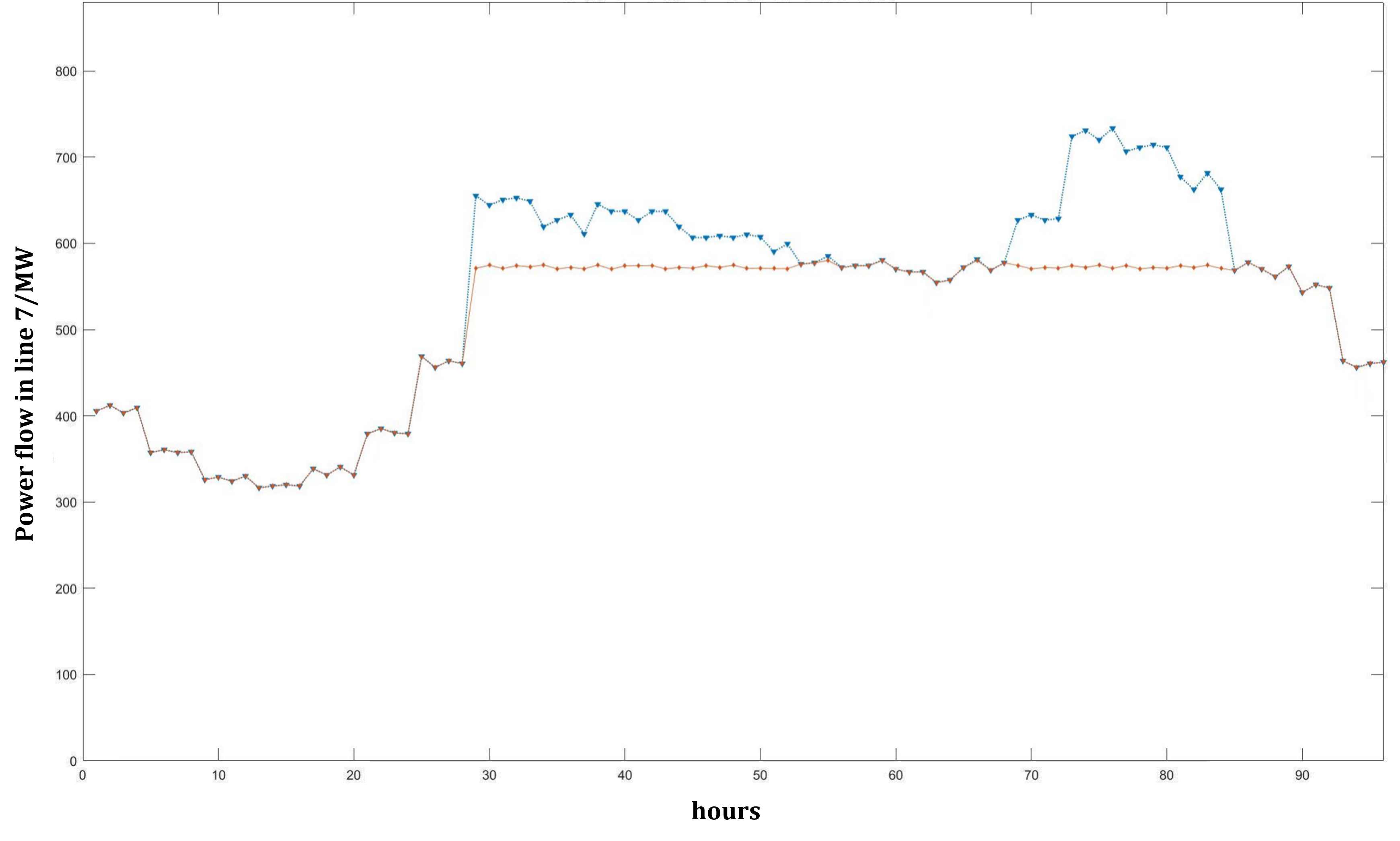} 
  \caption{Result of congestion management with capacity boundary of $580$MW} 
  \label{fig:Gen_PR1} 
\end{figure}

From Table \ref{table:1} we know that transmission line $7$ connects bus $8$ to bus $9$. When encountering a congestion in this line, the first instinct is to adjust the output of Gen.$4$ because it is located on bus $9$. Although selecting Gen.$4$ as the target generator may release the congestion on this transmission line, it will also sacrifice the economic advantage of Gen$4$. In order to select the most congestion-relief-effective target generator, we compare the value of generalized GSDF, especially between $GSDF_{G_4}$ and $GSDF_{G_5}$. Results show that Gen.$4$ is the most effective target generator.

The selection of balancing generator is also vital. In order to transfer the congested power to a distant area, which can naturally avoid congestion in other transmission lines near line $7$, the balancing generator always keep fine distance away from the target generator. Under such condition, generator with the most significant value of GSDF should be selected as balancing generator.

Note that the definition of electric distance between node $i$ and $j$ is given as:
\begin{align}
  &D_{ij}=Z_{ij,eq}=(Z{ii}-Z{ij})-(Z{ij}-Z{jj}) \label{NCED:volatility}
\end{align}
where $Z{ij}$ is the element of system node impedance matrix on line $i$ column $j$.

We calculate the sensitivity then apply the result to congestion management. In a 24-hour simulation, the power flow observations of line $7$ are shown in Fig~\ref{fig:Gen_PR1}. The upper bound is set to be 580 MW.

We further define volatility of the congestion management result as follows:
\begin{align}
  &vol=\frac{1}{\sum_{t}S_t}\sum_{t=1}^{64} (\frac{T_k^{1,t}}{T_k^{0,t}}-1)\times S_t\times 100\% \label{NCED:volatility}
\end{align}
where $S_t$ is a binary variable to show the congestion status. Here $S_t=1$ if and only if any congestion happens.


One may find that the management strategy tends to restrict the maximal output steady around the upper bound. The fluctuation of the result is around 10 MW and the volatility of the targeting power flow $vol_{580MW}$ is just below 2\%. This indicates that the proposed GSDF is accurate. 

More importantly, the fluctuation of power flow in line~7 is unidirectional, meaning that the power flow is strictly below the upper boundary which is set to 580~MW. With one transmission line congestion-managed, re-simulation should be operated in case of congestion in other transmission lines. According to the strategy mentioned in Fig.~\ref{fig:GSDF_trans}, the congestion operation loop should exit until all transmission lines satisfies constraint~\eqref{NCED:cons_Tlimits}.

\begin{figure}
  \centering
  \includegraphics[width=0.48\textwidth]{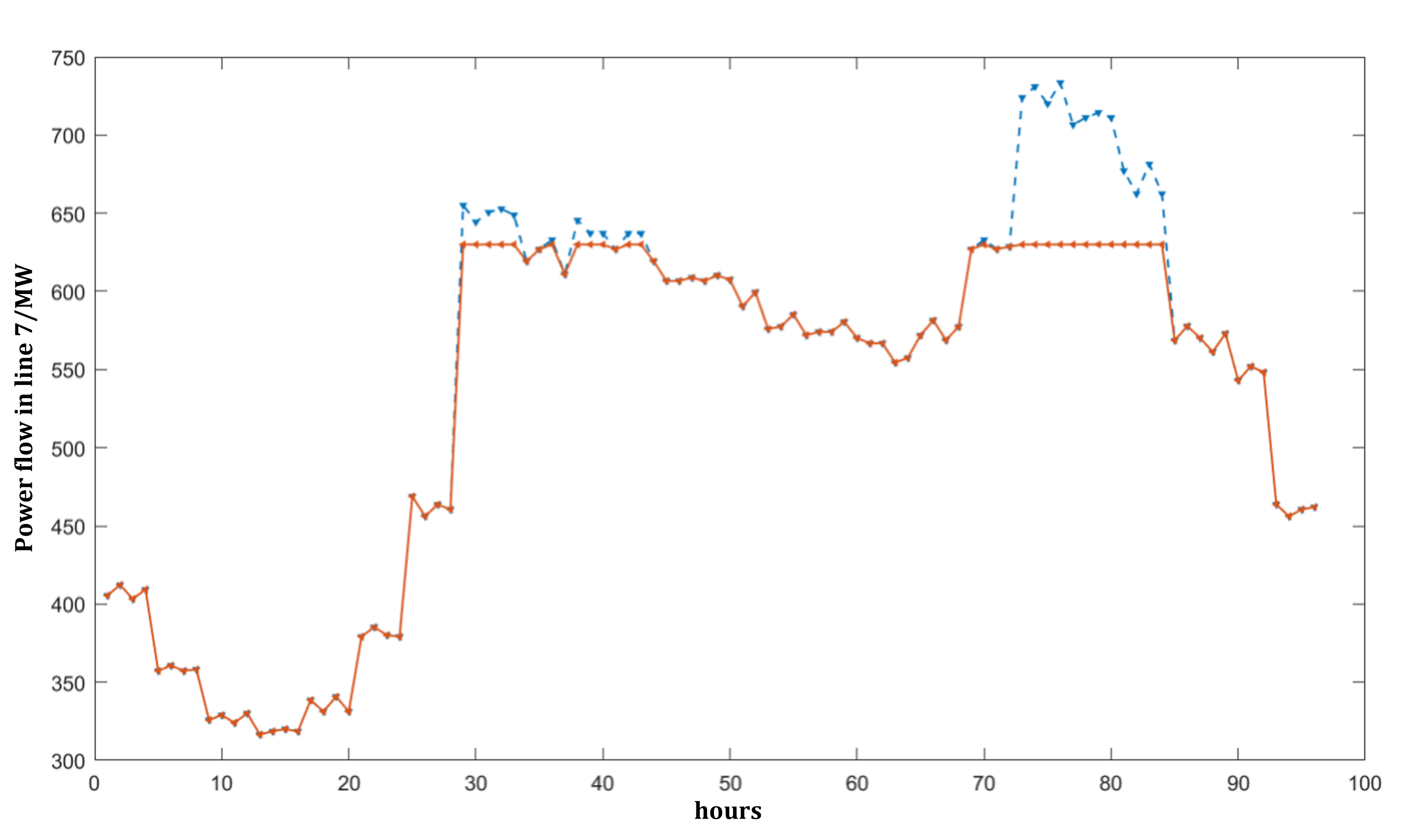} 
  \caption{Result of congestion management with capacity boundary of 630~MW} 
  \label{fig:Gen_PR2} 
\end{figure}

When changing the capacity bound to 630~MW, the fluctuation of the result is around 5~MW. The result is shown in Fig~\ref{fig:Gen_PR2}. Compared to Fig~\ref{fig:Gen_PR1}, the result of congestion management with loose boundary subjects to fewer transmission shedding. The hours of congestion management in Fig~\ref{fig:Gen_PR2} also decreases compared to the result with strict constraints.

The fluctuation of the targeting power flow $vol_{630MW}$ is below 1\%. This indicates that the fluctuation tends to decrease when the congestion is less severe. Still, the power flow is strictly below the upper boundary. The result explains the positive relevance between congestion management accuracy and the strictness of the transmission line power flow boundary. 

\section{Conclusion}

In this paper, we proposed an efficient transmission congestion management approach using linearized power flow. The proposed approach could consider voltage amplitude and reactive power flow while maintaining the linearity. Also, this approach includes target-balancing generalized GSDF and ensures the computational efficiency of GSDF estimation. Results indicate that the proposed congestion management strategy is valid.

The superior precision performance and simplified linearization expressions make our method competitive in practical applications when compared with other options based on DC or AC power flow.

As for future work, we plan to consider more complex scenarios, such as AC/DC hybrid systems or inverter-based low-inertia systems. 

\bibliographystyle{IEEEtran}
\bibliography{references}

\begin{thebibliography}{10}
\providecommand{\url}[1]{#1}
\csname url@samestyle\endcsname
\providecommand{\newblock}{\relax}
\providecommand{\bibinfo}[2]{#2}
\providecommand{\BIBentrySTDinterwordspacing}{\spaceskip=0pt\relax}
\providecommand{\BIBentryALTinterwordstretchfactor}{4}
\providecommand{\BIBentryALTinterwordspacing}{\spaceskip=\fontdimen2\font plus
\BIBentryALTinterwordstretchfactor\fontdimen3\font minus
  \fontdimen4\font\relax}
\providecommand{\BIBforeignlanguage}[2]{{%
\expandafter\ifx\csname l@#1\endcsname\relax
\typeout{** WARNING: IEEEtran.bst: No hyphenation pattern has been}%
\typeout{** loaded for the language `#1'. Using the pattern for}%
\typeout{** the default language instead.}%
\else
\language=\csname l@#1\endcsname
\fi
#2}}
\providecommand{\BIBdecl}{\relax}
\BIBdecl

\bibitem{4384982}
C.~Chompoo-inwai, C.~Yingvivatanapong, P.~Fuangfoo, and W.-J. Lee,
  ``Transmission congestion management during transition period of electricity
  deregulation in thailand,'' \emph{IEEE Transactions on Industry
  Applications}, vol.~43, no.~6, pp. 1483--1490, 2007.

\bibitem{2018ch}
Z.~Tan, G.~Ruan, H.~Zhong, and Q.~Xia, ``Security pre-check method of bilateral
  trading adapted to independence of power exchange (in chinese),''
  \emph{Automation of Electric Power Systems}, vol.~42, no.~10, pp. 106--113,
  2018.

\bibitem{7407415}
M.~Mahmoudian~Esfahani and G.~R. Yousefi, ``Real time congestion management in
  power systems considering quasi-dynamic thermal rating and congestion
  clearing time,'' \emph{IEEE Transactions on Industrial Informatics}, vol.~12,
  no.~2, pp. 745--754, 2016.

\bibitem{1709109}
A.~Kumar, S.~Singh, and L.~Lai, ``Impact of tcpar on congestion clusters and
  congestion management using mixed integer nonlinear programming approach,''
  in \emph{2006 IEEE Power Engineering Society General Meeting}, 2006, pp. 7
  pp.--.

\bibitem{9594777}
C.~A. Ordóñez~M, A.~Gómez-Expósito, G.~E. Vinasco~M, and J.~M. Maza-Ortega,
  ``Optimal coordinated operation of distributed static series compensators for
  wide-area network congestion relief,'' \emph{Journal of Modern Power Systems
  and Clean Energy}, vol.~10, no.~5, pp. 1374--1384, 2022.

\bibitem{gaonkar_nanannavar_manjunatha_2017}
V.~Gaonkar, R.~B. Nanannavar, and Manjunatha, \emph{Power system congestion
  management using sensitivity analysis and particle swarm optimization}, 2017.

\bibitem{4762626}
M.~H. Moradi, S.~Dehghan, and H.~Faridi, ``Improving zonal congestion relief
  management using economical \& technical factors of the demand side,'' in
  \emph{2008 IEEE 2nd International Power and Energy Conference}, 2008, pp.
  1027--1032.

\bibitem{7079504}
H.~Khani, M.~R. Dadash~Zadeh, and A.~H. Hajimiragha, ``Transmission congestion
  relief using privately owned large-scale energy storage systems in a
  competitive electricity market,'' \emph{IEEE Transactions on Power Systems},
  vol.~31, no.~2, pp. 1449--1458, 2016.

\bibitem{2020con}
G.~Ruan, H.~Zhong, Q.~Xia, C.~Kang, Q.~Wang, and X.~Cao, ``Integrating
  heterogeneous demand response into {N}-1 security assessment by
  multi-parametric programming,'' in \emph{2020 IEEE PES Innovative Smart Grid
  Technologies Conference (ISGT)}.\hskip 1em plus 0.5em minus 0.4em\relax IEEE,
  2020, pp. 1--5.

\bibitem{8468081}
Z.~Yang, K.~Xie, J.~Yu, H.~Zhong, N.~Zhang, and Q.~Xia, ``A general formulation
  of linear power flow models: Basic theory and error analysis,'' \emph{IEEE
  Transactions on Power Systems}, vol.~34, no.~2, pp. 1315--1324, 2019.

\bibitem{7954975}
Z.~Yang, H.~Zhong, A.~Bose, T.~Zheng, Q.~Xia, and C.~Kang, ``A linearized opf
  model with reactive power and voltage magnitude: A pathway to improve the
  mw-only dc opf,'' \emph{IEEE Transactions on Power Systems}, vol.~33, no.~2,
  pp. 1734--1745, 2018.

\end{thebibliography}

\end{document}